# Confirmation of the Electrostatic Self-Assembly of Nanodiamonds

**Lan-Yun Chang,[a] Eiji Ōsawa[b] and Amanda S. Barnard*,[c]**

A reliable explanation for the underlying mechanism responsible for the persistent aggregation and self-assembly of colloidal 5 nm diamond nanoparticles is critical to the development of nanodiamond-based technologies. Although a number of mechanisms have been proposed, validation has been hindered by the inherent difficulty associated with the identification and characterisation of the inter-particle interfaces. In this paper we present results of high resolution aberration corrected electron microscopy and complementary computer simulations to explicate the features involved, and confirm the electrostatic interaction mechanism as the most probable cause for the formation of agglutinates and agglomerates of primary particles.

Diamond nanoparticles are emerging as an important material for a variety of high performance technologies [1]. In particular, their optical [2] and chemical properties [3,4], coupled with their lack of cytotoxicity [5], make them a preferred candidate for biomedical applications [6,7]. This includes drug delivery in the treatment of cancer [6], heart disease [8] and in regenerative medicine [9]. It also includes biolabelling [10] based on their fluorescence in the visible range [2], owing to a number of photoactive point defects [11] that exhibit a high quantum yield [12], superior photostability [13] and structural stability [14]. Nanodiamonds are also being tested as components in the next generation of ultrasensitive contrast agents for magnetic resonance imaging [15]. In each case however, the realisation of technologies based on their properties is contingent on our ability to control nanodiamonds on an individual and collective basis. The formation of robust porous microstructures is highly desirable for drug delivery, whereas effective biomarkers must be discrete and as small as possible.

For many years, detonation nanodiamond, also referred to as *ultrananodiamonds* (UND), at around ~4.8 nm in size, appeared set to lead in this endeavour, but development was hindered by an inability to separate samples into individual single-digit nanoparticles [16]. These ultra-small nanodiamonds exhibit a very strong tendency to aggregate, and form robust superstructures on the order of ~100 nm in size [17]. For many years it was speculated that the aggregation was due to van der Waals interactions, but this hypothesis was dismissed when it was found that the agglomerates strongly resisted dispersion [18]. It was then argued that chemical bonds between the individual particles were responsible, but this suggestion was inconsistent with the XRD results showing that the diamond nanoparticles remain discrete. More recently, in attempts to understand the nature of agglomerates, researchers discovered that (given suitable conditions) diamond nanoparticles spontaneously self-assembled into ordered structures [19]. Based on these observations the terms agglomerate and agglutinate were introduced to distinguish between (apparently) random collections of diamond nanoparticles and the ordered superstructures, but the underlying mechanism for the formation of agglutinates remained elusive [16].

The mechanism responsible for the strong but long ranged interactions between colloidal diamond nanoparticles came in 2007 [20], when it was shown that individual particles exhibit anisotropic facet-dependent variations in the surface electrostatic potential. The (100) facets exhibited a positive net charge, and the (111) were either found to be charge neutral or to exhibit a net negative charge, depending upon whether surface graphitization was efficient. Given that diamond nanoparticles are polyhedral; this gives rise to a multi-pole that determined how nanodiamonds interact with each other [21] and with other molecules [22].

It was then shown that, due to the potential energy difference between different crystallographic orientations, preferred particle-particle orientations existed. Two thermodynamically favourable inter-particle configurations were identified, with positive $(100)^+$ facets interfacing with near-neutral $(111)^0$ facets, or negative $(111)^-$ facets interfacing with near-neutral $(111)^0$ facets. In both cases a neutral facet was an essential participant, and these interactions were termed coherent interfacial Coulombic interactions (CICI).

Alternative configurations arising from random electrostatic interactions were termed incoherent interfacial Coulombic interactions (IICI). These are both distinct from random weak interactions such as van der Waals forces that generally result in loosely aggregated powders. Depending upon whether CICI or IICI interactions dominated, nanodiamonds were predicted to form agglutinates or agglomerates, as described in table 1. Until now however, this prediction has not been experimentally corroborated, partly due to the inherent difficulty associated with extracting quantitative or visual information on the relative amounts of IICI and CICI within a given sample, and partly due to the complexity of the reaction environment. There are a number of dependent physical variables influencing the mesostructure of samples during characterization, such as temperature and pressure induced surface restructuring, forced evaporation of the surrounding medium, space charging and primary particle mobility. Once the existence of CICI and IICI are confirmed, more detailed studies of how these mechanisms may be used to engineer diamond nanoparticles samples may begin.

[a] *Monash Centre for Electron Microscopy and School of Chemistry, Monash University, Clayton, VIC, Australia.*
[b] *NanoCarbon Research Institute, AREC, Faculty of Textile Science and Technology, Shinshu University, Ueda, Nagano, Japan*
[b] *CSIRO Materials Science and Engineering, Clayton, Australia. Fax: +61 3 9545 2059; Tel: +61 3 9545 7840;*
** E-mail: amanda.barnard@csiro.au*

† Electronic Supplementary Information (ESI) available: Details of the sample preparation and computational methodology. See http://dx.doi.org/10.1039/b000000x/

**Table 1** Hierarchical nanodiamond aggregation.

| Name | Size (nm) | Interaction Mechanism | | | |
|---|---|---|---|---|---|
| | | Type | Nature | Configuration | Terminology |
| Primary Particle | 4.7 | - | - | - | Isolated |
| Agglutinate | c.a. 60 | Interfacial | Electrostatic | Ordered | CICI [a] |
| Agglomerate | 100~200 | Interfacial | Electrostatic | Random | IICI [b] |
| Powder | >1000 | Intergranular | van der Waals | Random | VDWA [c] |

[a] Coherent Interfacial Coulombic Interaction, [b] Incoherent Interfacial Coulombic Interaction, [c] van der Waals Aggregation.

In order to definitely confirm the preferred orientations for the strong and long-ranged, nanodiamonds-nanodiamond interactions, a number of observations are required. Firstly, that the facet-facet separation distance is on the order of 0.19 nm, as predicred by the simulations (and described in reference 21). Secondly, that the particle-particle interfaces are dominated by negative/neutral or positive/neutral interactions as predicted by the computational study [21]. If this is found to be the case, this will be evidenced by the orientation of interfacing particles, and it is expected that one would observe linear superstructures such as chains and loops, depending on the relative prevalence of the different types of preferred interactions [21].

In an attempt to identify either (or both) of these signatures, we have undertaken high-resolution transmission electron microscopy (HRTEM) of colloidal diamond nanoparticles. The dispersed nanodiamond particles are now routinely prepared at a rate of about 2kg/month by a six-step procedure in Ueda, Japan [3], in which the major step is beads-milling with 30µm zirconia beads, as described in the Electonic Supplementary Information (ESI).

Conventional HRTEM operating at medium voltage (100 - 300kV) has been used in the past to characterise the structure of diamond nanoparticles [23,24]. However, due to the weak scattering power of single particles, and the effect of delocalisation due to aberrations introduced by the objective lens, interpretation of such HRTEM images has been difficult. Here, we use aberration-corrected TEM operated at 80kV to image the diamond nanoparticles, particularly in an attempt to identify their interfacial structures. The advantage of using such an instrumental setup is that the aberration-corrector greatly reduces the delocalisation in HRTEM images [24], and operating at 80kV minimises the radiation damage to the sample, and increases the image contrast due to a stronger effective interaction at such accelerating voltage [25]. Nevertheless, it is still expected that TEM specimen preparation and exposing the sample under electron beam

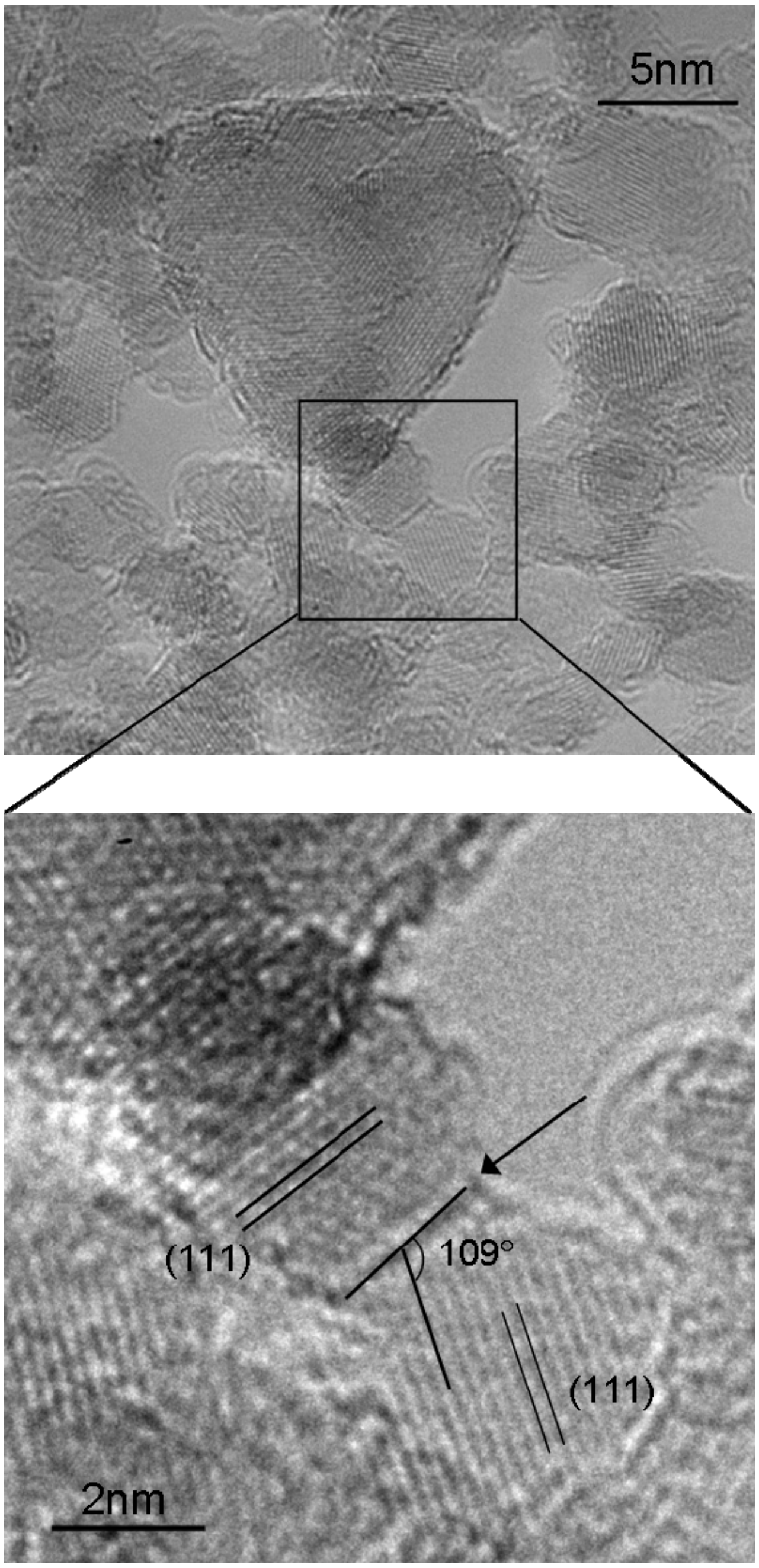

**Fig.1**. The arrow indicates (111)|(111) interface between two 4 nm sized nanodiamonds. The orientation and the separation of the (111) lattice fringes are also shown.

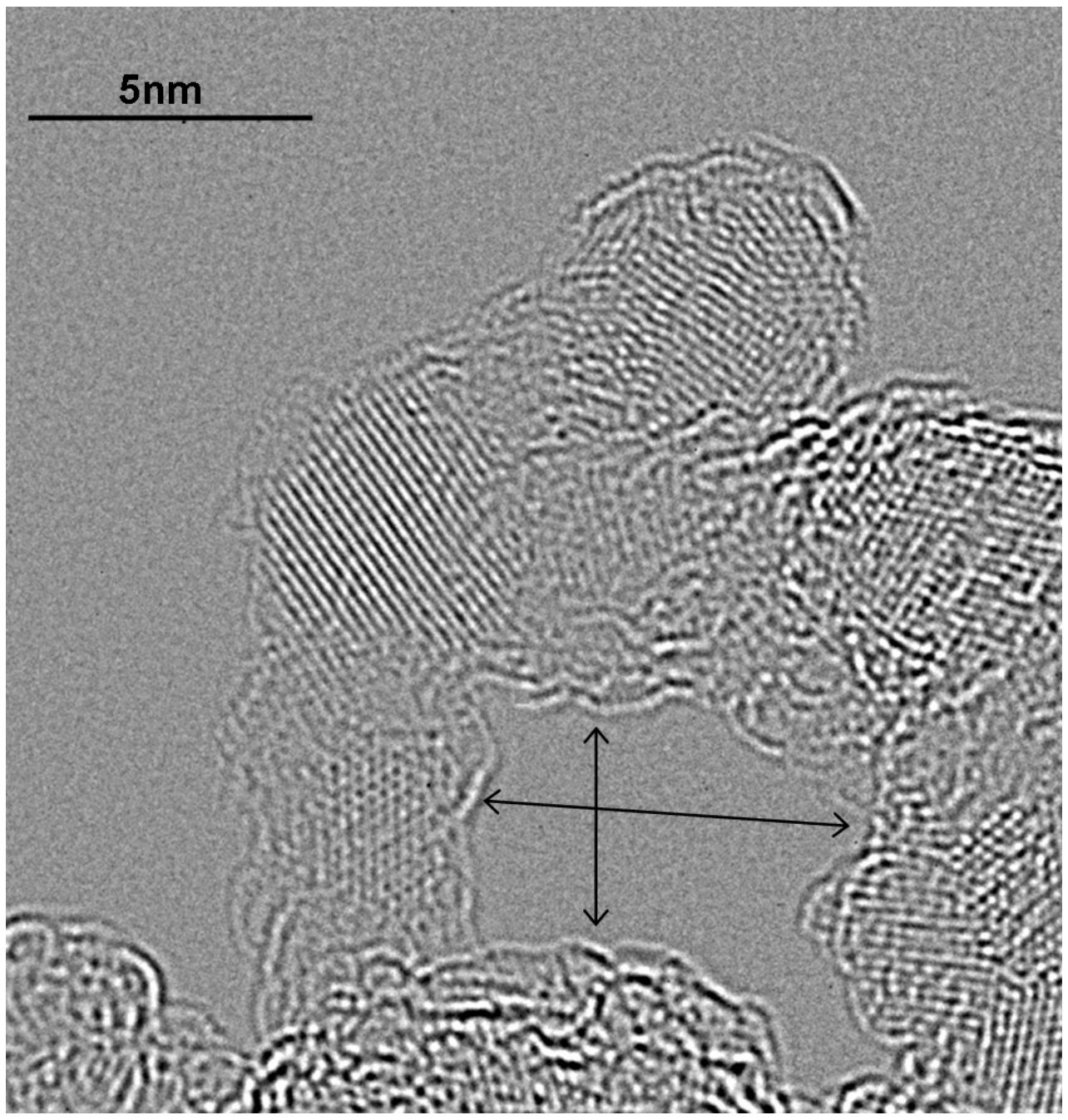

**Fig.2**. A three-dimensional (3-D) loop of nanodiamonds. The size of the individual nanodiamond ranges from 4-5nm, and the size of void within the loop is 4 nm (vertical width) by 6 nm (horizontal width). Due to the 3-D configuration, the locations of the interfaces are obscured.

during imaging may have an impact on the sample, for the reasons highlighted above.

Even with this state-of-the-art imaging instrumentation, characterisation of agglutinates and agglomerates is still challenging, due to the three-dimensional nature of the aggregated superstructures, the co-existence of CICI and IICI within the one sample, and the existence of more than one allotrope. In most cases, these particles are decorated with a partial coating of fullerenic $sp^2$ carbon around the diamond-like $sp^3$ core; a structure which is known as a bucky-diamond [26]. This is not unexpected, and is generally consistent with previous experimental observations [23], computational simulations [20] and theoretical predictions [27]. After careful preparation and patient examination, we were able to achieve sufficient dispersion and identify numerous regions of the sample only one particle layer in thickness. Examples of these results are shown in figures 1, 2 and 3, and a the details of the Fourier transform analysis are provided in the Electronic Supplementary Information (ESI).

In figure 1 we see an interface between two 4 nm diamond nanoparticles, with the {111} lattice planes clearly visible. We have indexed this interface, based on the orientation of the two participating particles and the angle between the interface and the {111} planes, and confirm that it is (111)|(111), which is consistent with the prediction that an interface between $(111)^-$ facets and near-neutral $(111)^0$ facets will be stable. We can also see that this interface contains a layer of fullerenic carbon (darker contrast at this imaging condition), which is also consistent with the prediction that only one surface (at the interface) would be graphitized and negatively charged. Unfortunately, the facet-facet separation distance can not be identified unambiguously in the case, due to the off-zone axis

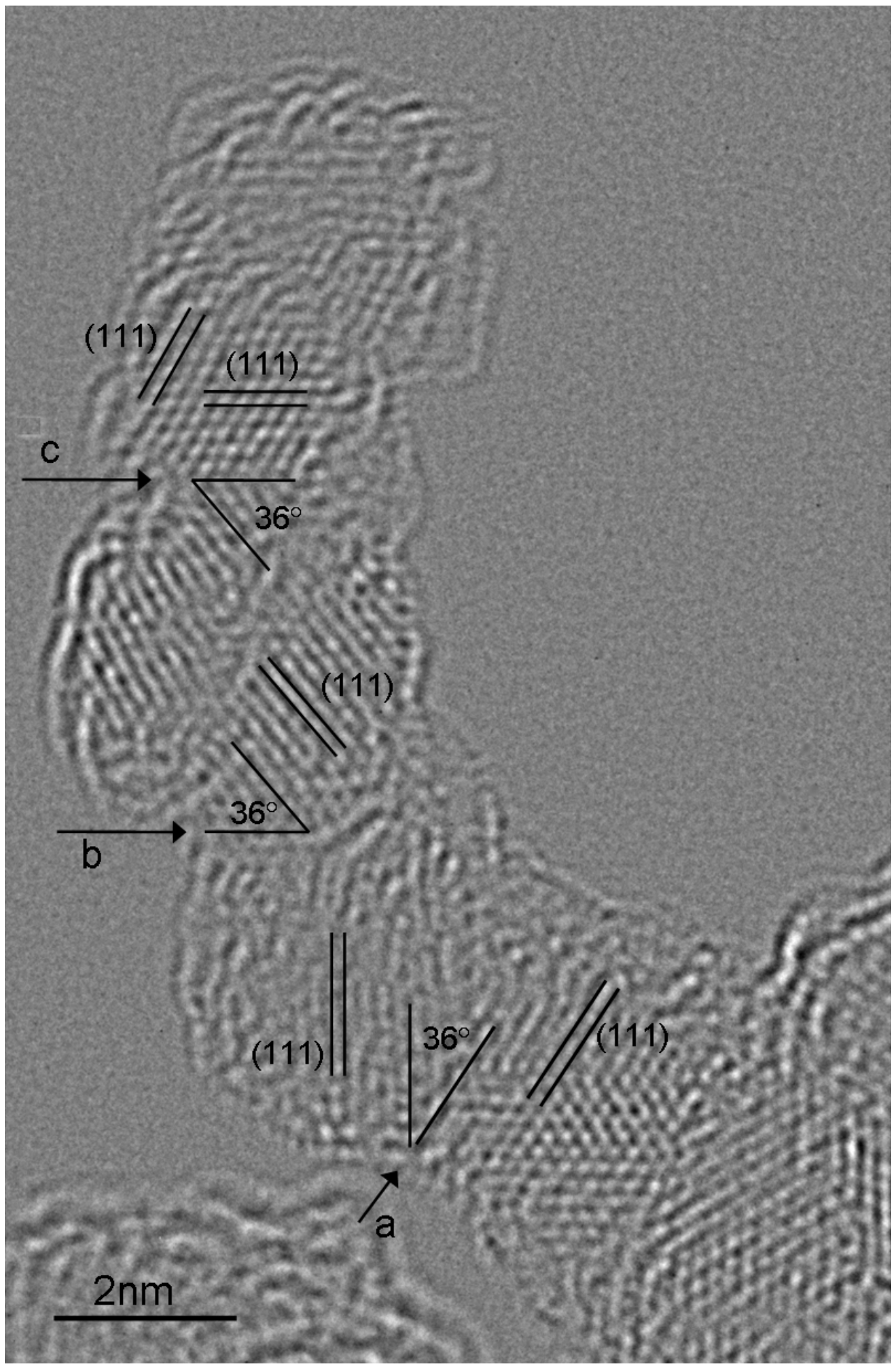

**Fig.3**. A linear chain of four ~4.5nm sized nanodiamonds. Their interfaces are determined to be (111)|(220) type (labeled as a,b,c), with one interface that can not be uniquely determined.

orientations of the particles and slight delocalisation arising from the residual aberrations.

In figure 2 we see a loop of 4 nm diamond nanoparticles surrounding a 4-6 nm void. In this case, due to the three-dimensional nature of the superstructure, the precise locations of the interfaces are obscured, and characterization of the interfacial orientations is not possible. However, it was predicted that agglutinates would contain voids [21]. The limitations inherent in CICI self-assembly manifest as a tiling problem, and to date no space filling two dimensional solutions have been identified that do not contain a relatively high fraction of voids. Recent results from nitrogen adsorption isotherms of the dried powder demonstrated the presence of nano-voids, with a characteristic size and volume which matched the observed quantity of nanophase water identified using differential scanning calorimetry (DSC) [28]. The size of the voids identified during our study is consistent with these measurements [28] and those predicted [21].

Figure 3 presents a chain of four detonation nanodiamonds. The orientation of the nanodiamonds with respect to one another is determined by analysing the angles between the interface and the {111} lattice fringes, and the orientation of the particles as determined by the Fourier transforms of the sub-regions of individual nanodiamond. The location of three interfaces is highlighted by the arrows (marked a, b and c), and the fourth interface cannot be uniquely determined. We see here that the interfaces of these nanodiamonds are not of the $(111)^-|(111)^0$ type as shown in figure 1, but are a combination of {111} and {110} facets (formally {220}).

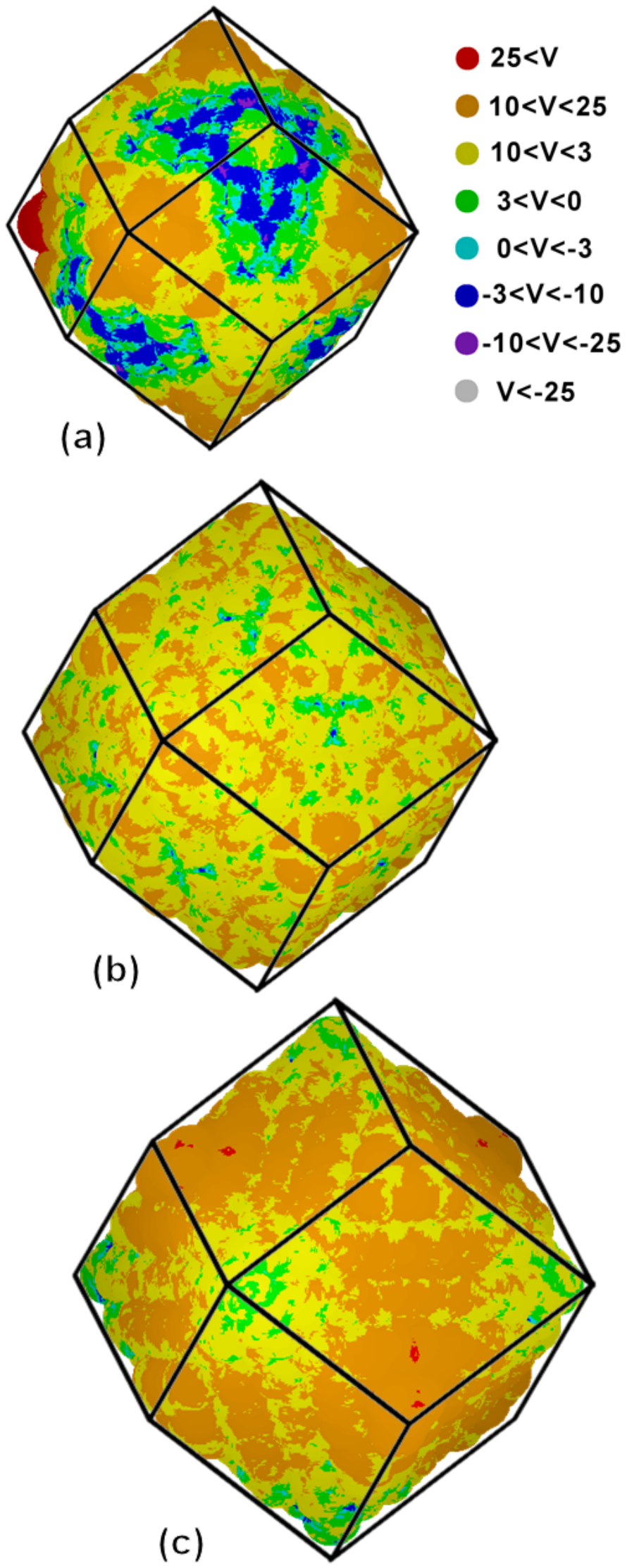

**Fig.4**. DFTB simulations of the surface electrostatic potential of dodecahedral diamond nanoparticles containing (a) 844 atoms, (b) 1232 atoms, and (c) 1722 atoms, measuring 2.2 nm, 2.5 nm and 2.9 nm, respectively.

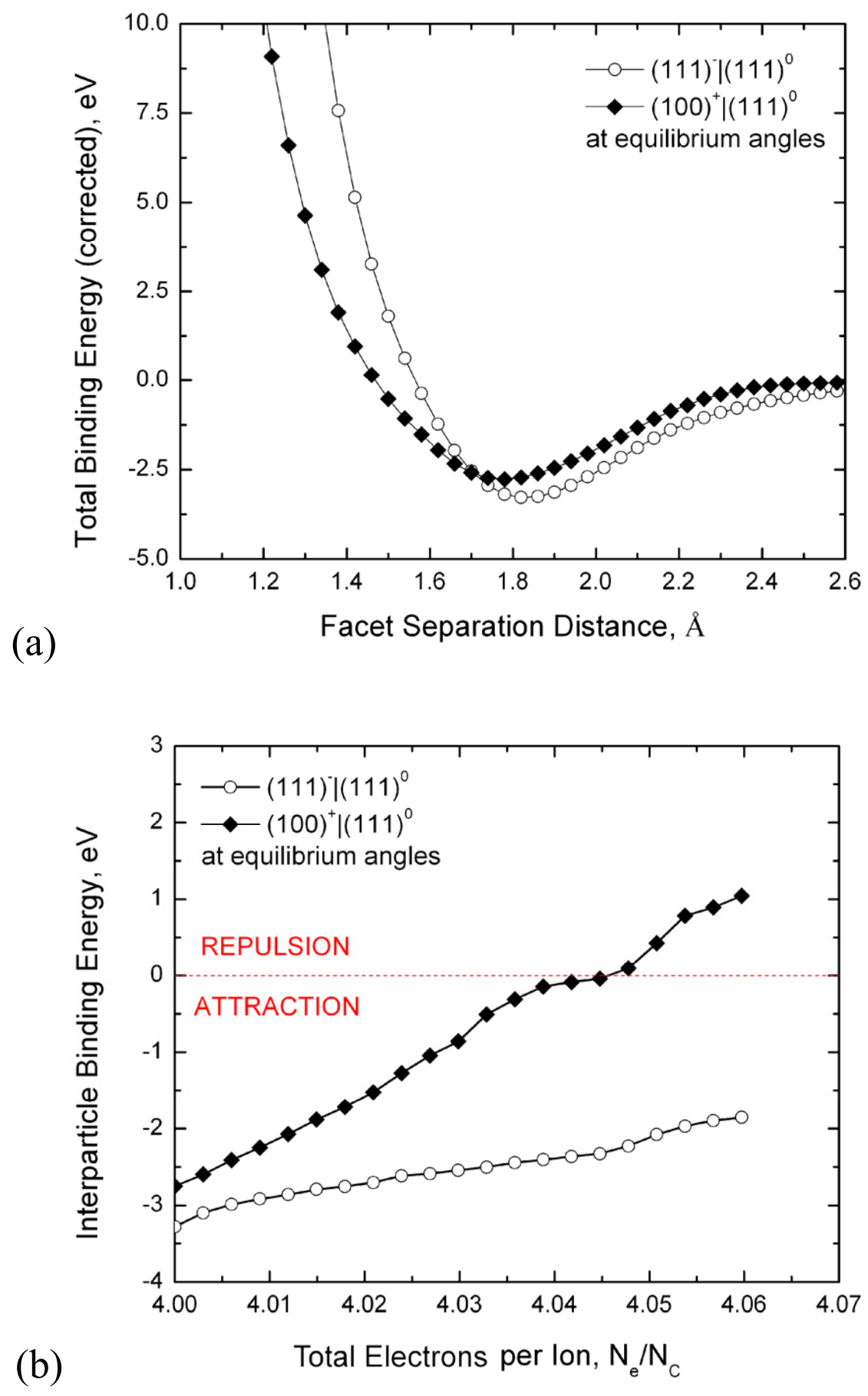

**Fig.5**. (a) The potential energy well due to CICI between nanodiamonds calculated using DFTB simulations, and (b) the effect of injected excess charge on the stability of $(111)^0|(111)^-$ or $(100)^+|(111)^0$ interfaces; where $N_e$ is the total number of electrons in the system, and $N_c$ is the total number of carbon atoms. Increasing the number of excess electrons (anionic charges) over these pairs the satisfaction of the net positive potential on the $(100)^+$ facets as the system recovers surface charge neutrality. This leads to weakening of the electrostatic $(100)^+|$ $(111)^0$ interactions and eventually to separation of this nanodiamond pair. The term "equilibrium angle" referes to the lowest energy angle of rotation about the facet-facet normal.

Interfaces containing {110} facets have not been considered before, and since a (111) facet is also involved, they may be negative, near-neutral or positive and still satisfy the selection rules (above). The surface electrostatic potential of (110) facet was briefly investigated in reference [20]. Although the results suggested that the facets were charge neutral on average, the cuboid morphologies were experimentally unrealistic, so the assignment of charge was inconclusive.

We therefore undertaken density functional tight binding (DFTB) simulations on three rhombic dodecahedral nanodiamond structures, enclosed entirely by {110} facets, using the same computational methods as described in references [20] and [21]. These results are presented in figure 4, where we can see that the surface facets are indeed near-neutral, with a slight positive charge. This suggests that (in order to obey the selection rules) the interfaces observed experimentally are due to the CICI mechanism, and involve negative charged $(111)^-$ facets that are fully graphitized, and near-neutral $(110)^0$ facets.

One issue that emerges from these observations and simulations, is the lack of $(100)^+|(111)^0$ interfaces, that were predicted to be almost as energetically favourable as the $(111)^-|(111)^0$ interactions. A more extensive search failed to find evidence of $(100)^+$ facets interfacing with near-neutral $(111)^0$ facets at all. To understand why this is the case we must first think about what is happening to the nanodiamonds during imaging. Although the graphitized surfaces are conducting, and will allow for the excess charge deposited by the electron beam to transport, the diamond cores are insulating, and many of the surfaces have not efficiently graphitized. This means that much of the charge deposited on the particles will contribute to reside there, and will logically relocate to the $(100)^+$ facets first, to balance the net positive potential and recover neutrality. Once all of the $\{100\}^+$ sites have been satisfied, the incident electrons will begin to decorate the near-neutral $(111)^0$ and $(110)^0$ facets. Therefore, artificial anionic charging at doses consistent with imaging would disrupt the CICI mechanism, converting $(100)^+|(111)^0$ interfaces to $(100)^0|(111)^0$ interfaces subject to weak van der Waals forces.

To test this hypothesis we have undertaken a series of DFTB simulations investigating the effect of excess charge on pairs of interacting diamond nanoparticles. Shown in figure 5a are the potential energy wells for $(100)^+|(111)^0$ and $(111)^-|(111)^0$ type interfaces at the lowest energy inter-facet angle of rotation. By taking the pairs of particles at the configuration corresponding to the minima of each well, and systematically injecting electrons (where $N_e$ is the total number of electrons in the system), we can see the charge-dependent change in the well-depth in figure 5b. In the case of the $(100)^+|(111)^0$ interface, the interaction becomes rapidly more unstable with increasing anionic charge, and eventually becomes a forbidden interaction at 0.045 extra electrons per carbon atom (where $N_c$ is the total number of carbon atoms in the system). Experimentally, an electron dose rate of approximately 2000 electrons per $Å^2$ per second was used for HRTEM imaging. For an area where a single 4 nm single particle falls within the incident electron illumination, this dose rate is translated into approximately 600 electrons per atom per second. Assuming there is no charge dissipation, or transport to other diamond nanoparticles, a dose sufficient to separate a $(100)^+|(111)^0$ interface is achieved in approximately 75 microseconds, thereby explaining why this interface is not observed.

These results have another interesting consequence, being the prospect of controlling these interactions with electronic charge. This has been shown before using variations in pH [19], which is also important in drug delivery [7], but also introduces the opportunity to disperse diamond nanoparticles samples without the need for mechanical milling with zirconia beads [18].

The confirmation of the CICI interactions, and the possibility of control them using charge, also suggests that substrate patterning may also be possible, which would be a great advance in the field of MEMS and NEMS [29]. Currently diamond nanoparticles are an indispensible ingredient in the seeding of ultrananodiamond thin films [30], and it has already been demonstrated that self-assembled diamond nanoparticles may act as a template for subsequent film growth via chemical vapour deposition [31]. Electrostatic dispersion and assembly of diamond nanoparticles on a substrate has already been demonstrated [32-34], and one can see why the tantilising possibility of controlled pre-patterning of substrate features on the order of 6–5 nm would be a revolution.

Therefore, future work will be directed toward processing methods aimed at exercising more control over the shape of individual nanodiamonds, and understanding how difference facet-dependent surface coatings and functionalization will affect the results, and whether it can also be used to deliver greater specificity.

## Acknowledgements

Computational resources for this project have been supplied by the National Computing Infrastructure (NCI) national facility under Partner Grant q27. Aberration-corrected Titan 80-300 (FEI Company) used in this project has been funded by Australian Research Council Grant LE0454166.

## Notes and references

 

# Electronic Supplementary Information: Confirmation of the Electrostatic Self-Assembly of Nanodiamonds

**Lan-Yun Chang,[a] Eiji Ōsawa[b] and Amanda S. Barnard*,[c]**

## Experimental

### Preliminary screening

Not all the crude agglutinates are crushable into primary particles by beads-milling. The most reliable criterion for the crushability of crude detonation product is zeta-potential in aqueous suspension, which must be higher than +50mV or lower than -50mV as determined on a Zetasizer Nano-ZS (Malvern Instruments, England). We use fluffy, grey-coloured fine powder made by Guangzhou Panyu Guangda Electromechanical Company (Guangzhou, China).

### Slurry preparation

Suspension of 10% agglutinate powder in water was prepared by agitating the mixture with a T.K.Robomix (model f, Primix Co., Tokyo) at a rotor speed of 5000 rpm for 30 min, followed by immersion into a powerful supersonic washing bath (model W-113 MK-II, Honda Electronics Co., Toyohashi) for at least two hours. At the end of this period, thick, homogeneous and light grey slurry was obtained, which soon produces dense precipitates.

### Beads-milling

The slurry prepared above is charged into a slurry reservoir of a vertical Ultra Apex Mill (type UAM-015, manufactured by Kotobuki Industries Co., Tokyo), locally modified at the reservoir portion in order to perform milling under nitrogen atmosphere and to facilitate sampling, taking out the aliquot colloid and washing. Optimum conditions of milling were determined by using Experimental Design technique under the L9 orthogonal array formalism with the following variables: diameter of zirconia beads, packing rate of beads in the milling space, slurry concentration, rotor periphery speed, and number of circulation passes.

### Intense supersonic treatment with sonotrode

The collision between freshly dispersed diamond and abundant zirconia beads in the mill should be kept at a minimum in order to suppress the etching of zirconia by diamond. For this reason, the milling cycle was cut short of complete disintegration and the final step is taken care of by irradiation of intense ultrasonic wave from a 400W laboratory ultrasonic processor (type UP-400S manufactured by Hielscher Ultrasonic Co., Germany) attached with a titanium sonotrode (type H22).

[a] *Monash Centre for Electron Microscopy and School of Chemistry, Monash University, Clayton, VIC, Australia.*
[b] *NanoCarbon Research Institute, AREC, Faculty of Textile Science and Technology, Shinshu University, Ueda, Nagano, Japan*
[b] *CSIRO Materials Science and Engineering, Clayton, Australia. Fax: +61 3 9545 2059; Tel: +61 3 9545 7840;*
* *E-mail: amanda.barnard@csiro.au*

### Centrifugal separation of uncrushable residue

The sonicated colloidal solution was subjected to centrifugal separation of still remaining visible particles and other contaminants using a Table-top Centrifuge (type 5200 manufactured by KUBOTA Corporation) equipped with a swing bucket rotors (type ST-720, 200ml×4) at 3500 rpm for 1h. In average, 4.8 wt% of the used agglutinates were collected by this operation as uncrushable sediments. Supernatant liquids were collected by decantation and subjected to analysis.

### Particle-size distribution

The most useful criterion on the quality of colloidal solution is naturally the particle-size distribution. However the straightforward determination of particle size in nanocolloidal solution by means of dynamic light scattering method is unreliable and the reproducibility of results are in general unsatisfactory. We have long decided to seek a concentration range which gives the consistent size-distribution. In a typical run of a typical sample the size-distribution consists of a predominant and narrow distribution around 5nm and broad and minor peaks centred around 50-60 nm. The first peak comes from the primary particles of detonation nanodiamond but the origin of the second peak is still unknown. We used only those results where this typical pattern was obtained with reproducibility higher than 80%. In the particular sample used in this work, the particle-size distribution was determined by using three concentrations, 2.2, 2.0 and 1.8%. For each concentration, DLS measurements were repeated 440 times. The continuous series of measurements was divided into five equal sequences, and partial averages in each section are compared and the middle result with regard to the size of major peak was chosen as a representative distribution. Among the three representative results for three different concentrations, the middle one is finally chosen as the correct value. Hence the final answer is based on more than 1300 determinations, which takes about 45 min on a Particle Analyzer FPAR1000 equipped with an automatic sample FP3000 (manufactured by Ohtsuka Electronics Co., Tokyo). For the particular aqueous colloidal solution (2.0 w/v %) used in this work, distribution of the major peak was thus determined to appear at 4.1±0.5nm (99.4 wt %).

### Electron Microscopy

The analysis of the diamond nanoparticles samples was carried out using high resolution aberration-corrected transmission electron microscopy (HRTEM) operated at 80kV. In each case the images presented in the main text were supported by the Fourier transform analysis. For example, a sub-region in Figure 1 is shown in Figure S1c, where lattice image of a typical diamond nanoparticle is presented, shows strong {111} reflections and much weaker {200} reflections. Although {200} reflections are forbidden under kinematical conditions, it is well known that these "supposedly forbidden" {200} reflections are commonly observed in diamond due to double scattering [1]. In addition, the magnitude of {200} reflections in our case is much weaker than {111) reflections, which definitively confirms that the samples are members of the Fd3m space-group.

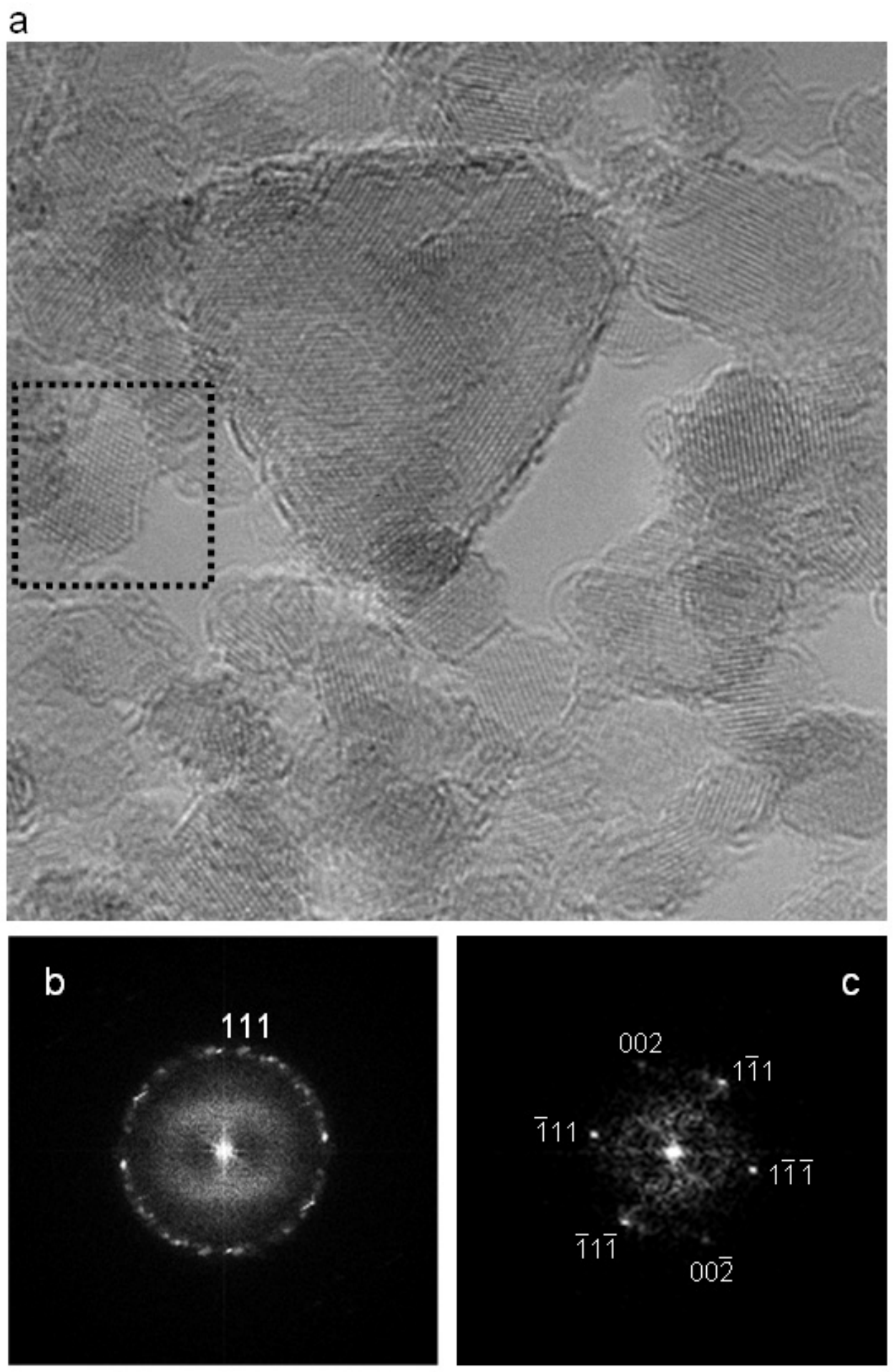

**Fig.S1. (a)** is the HRTEM image of a typical area of detonation nanodiamonds. **(b)** is the modulus of Fourier transform of figure **(a)**. The ring pattern shows that the nanodiamonds do not have preferred orientations. **(c)** is the modulus of Fourier transform of a sub-region shown in dashed square in **(a)**. The reflections were indexed and marked in **(c)**. The diamond nanoparticles in this sub-region is of [110] orientation and the weak {200} reflections, which should be forbidden in the kinematical conditions, are due to double scattering.

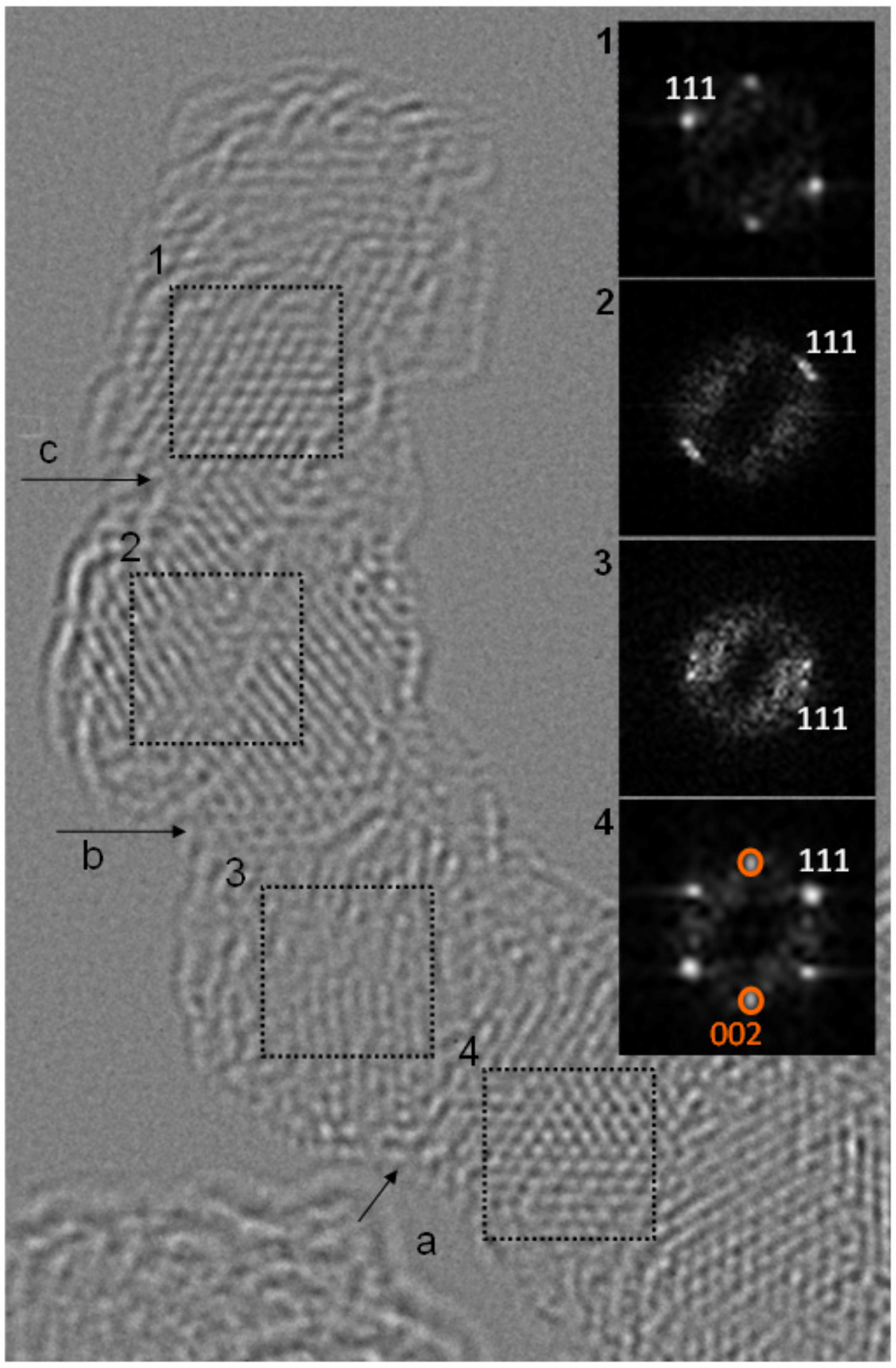

**Fig.S2.** HRTEM image of a linear chain and the corresponding moduli of Fourier transforms of individual diamond nanoparticles. The strong reflections in the Fourier transforms were measured to be {111} reflections. The Fourier transform of region 4 shows weak {002} reflections, which are due to double scattering.

The Fourier transforms of the areas of individual nanodiamond in Figure 3 in the main text, are shown in Figure S2. These results clearly show that the diamond nanoparticles in these samples consist of a diamond structured core with fullerenic surfaces, and are absent of defects. These Fourier transforms were used to determine the orientations of the nanodiamonds (which are in [110] orientations for region 1 and 4). The strong {111} reflections (also shown in the lattice fringes in all nanodiamonds in Figure 3 of the main text), along and the angles between {111} fringes and the interface, definitively determine the interfacial planes.

In addition to this, the modulus of Fourier transform of Figure 1a (a typical area of the detonation nanodiamond) shows a case where there is no preferred orientation. This highlights that this technique is not biased toward the assumption that all nanodiamonds self-assemble, and can also identify nanodiamonds in random orientations, as shown in Figure S1b.

## Computational

To be consistent with the results (and speculations) of references 2 and 3, the present study also uses the density functional based tight binding method with self-consistent charges (SCC-DFTB) [4,5]. In this approach, a universal short-range repulsive potential accounts for double counting terms in the Coulomb and exchange-correlation contributions as well as the inter-nuclear repulsion, and self-consistency is included at the level of Mulliken charges. Although not strictly an observable quantity, the Mulliken charges are not extracted post facto, and form an integral part of the energy functional which expresses local density fluctuations around a given atom. Mulliken charge fluctuations are calculated from the eigenvalue coefficients, and are algorithmically independent from bonding considerations and spatial partitioning schemes [6]. Therefore, although they generally have limited quantitative value, they are useful in illustrating bonding trends in heteronuclear systems, and near the surface of molecules and clusters such the nanodiamonds under consideration herein [7-11].

## Acknowledgements

Computational resources for this project have been supplied by the National Computing Infrastructure (NCI) national facility under Partner Grant q27. Aberration-corrected Titan 80-300 (FEI Company) used in this project has been funded by Australian Research Council Grant LE0454166.